\documentclass[aps,prl,showpacs,showkeys,amsmath,amssymb,
nofootinbib,
twocolumn,
floatfix,
superscriptaddress
]{revtex4-1}
\usepackage{graphicx}
\usepackage{times}
\usepackage{verbatim}
\usepackage{color}
\usepackage{cancel}
\usepackage[normalem]{ulem}

\usepackage{subfigure}
\usepackage{float}
\usepackage[toc,page]{appendix}
\usepackage{graphicx}
\usepackage{booktabs}
\usepackage{hyperref}
\usepackage{multirow}
\usepackage{enumitem}
\usepackage{soul}

\usepackage[none]{hyphenat}

\definecolor{ddedit}{rgb}{0,0.5,0}

\newcommand{\UF}{$\rm{^{235}U}$}
\newcommand{\PuN}{$\rm{^{239}Pu}$}
\newcommand{\UE}{$\rm{^{238}U}$}
\newcommand{\PuO}{$\rm{^{241}Pu}$}

\newcommand{\AveYieldMeas}{(5.89\pm0.07)\times10^{-43}~\rm{cm}^{2}/\rm{fission}}

\newcommand{\DblRatioMeas}{-0.300\pm0.024}

\newcommand{\AveYieldHM}{(6.18\pm0.04)\times10^{-43}~\rm{cm}^{2}/\rm{fission}}
\newcommand{\DblRatioHM}{-0.387\pm0.016}

\newcommand{\AveYieldHMM}{(6.18\pm0.16)\times10^{-43}~\rm{cm}^{2}/\rm{fission}}

\newcommand{\DblRatioHMM}{-0.387\pm0.018}

\begin{document}

\title{Improved Measurement of the Evolution of the Reactor Antineutrino Flux and Spectrum at Daya Bay}
\newcommand{\IHEP}{\affiliation{Institute~of~High~Energy~Physics, Beijing}}
\newcommand{\Wisconsin}{\affiliation{University~of~Wisconsin, Madison, Wisconsin 53706}}
\newcommand{\Yale}{\affiliation{Wright~Laboratory and Department~of~Physics, Yale~University, New~Haven, Connecticut 06520}} 
\newcommand{\BNL}{\affiliation{Brookhaven~National~Laboratory, Upton, New York 11973}}
\newcommand{\NTU}{\affiliation{Department of Physics, National~Taiwan~University, Taipei}}
\newcommand{\NUU}{\affiliation{National~United~University, Miao-Li}}
\newcommand{\Dubna}{\affiliation{Joint~Institute~for~Nuclear~Research, Dubna, Moscow~Region}}
\newcommand{\CalTech}{\affiliation{California~Institute~of~Technology, Pasadena, California 91125}}
\newcommand{\CUHK}{\affiliation{Chinese~University~of~Hong~Kong, Hong~Kong}}
\newcommand{\NCTU}{\affiliation{Institute~of~Physics, National~Chiao-Tung~University, Hsinchu}}
\newcommand{\NJU}{\affiliation{Nanjing~University, Nanjing}}
\newcommand{\TsingHua}{\affiliation{Department~of~Engineering~Physics, Tsinghua~University, Beijing}}
\newcommand{\SZU}{\affiliation{Shenzhen~University, Shenzhen}}
\newcommand{\NCEPU}{\affiliation{North~China~Electric~Power~University, Beijing}}
\newcommand{\Siena}{\affiliation{Siena~College, Loudonville, New York  12211}}
\newcommand{\IIT}{\affiliation{Department of Physics, Illinois~Institute~of~Technology, Chicago, Illinois  60616}}
\newcommand{\LBNL}{\affiliation{Lawrence~Berkeley~National~Laboratory, Berkeley, California 94720}}
\newcommand{\UIUC}{\affiliation{Department of Physics, University~of~Illinois~at~Urbana-Champaign, Urbana, Illinois 61801}}
\newcommand{\SJTU}{\affiliation{Department of Physics and Astronomy, Shanghai Jiao Tong University, Shanghai Laboratory for Particle Physics and Cosmology, Shanghai}}
\newcommand{\BNU}{\affiliation{Beijing~Normal~University, Beijing}}
\newcommand{\WM}{\affiliation{College~of~William~and~Mary, Williamsburg, Virginia  23187}}
\newcommand{\Princeton}{\affiliation{Joseph Henry Laboratories, Princeton~University, Princeton, New~Jersey 08544}}
\newcommand{\VirginiaTech}{\affiliation{Center for Neutrino Physics, Virginia~Tech, Blacksburg, Virginia  24061}}
\newcommand{\CIAE}{\affiliation{China~Institute~of~Atomic~Energy, Beijing}}
\newcommand{\SDU}{\affiliation{Shandong~University, Jinan}}
\newcommand{\NanKai}{\affiliation{School of Physics, Nankai~University, Tianjin}}
\newcommand{\UC}{\affiliation{Department of Physics, University~of~Cincinnati, Cincinnati, Ohio 45221}}
\newcommand{\DGUT}{\affiliation{Dongguan~University~of~Technology, Dongguan}}
\newcommand{\XJTU}{\affiliation{Department of Nuclear Science and Technology, School of Energy and Power Engineering, Xi'an Jiaotong University, Xi'an}}
\newcommand{\UCB}{\affiliation{Department of Physics, University~of~California, Berkeley, California  94720}}
\newcommand{\HKU}{\affiliation{Department of Physics, The~University~of~Hong~Kong, Pokfulam, Hong~Kong}}
\newcommand{\Charles}{\affiliation{Charles~University, Faculty~of~Mathematics~and~Physics, Prague}} 
\newcommand{\USTC}{\affiliation{University~of~Science~and~Technology~of~China, Hefei}}
\newcommand{\TempleUniversity}{\affiliation{Department~of~Physics, College~of~Science~and~Technology, Temple~University, Philadelphia, Pennsylvania  19122}}
\newcommand{\CGNPG}{\affiliation{China General Nuclear Power Group, Shenzhen}}
\newcommand{\NUDT}{\affiliation{College of Electronic Science and Engineering, National University of Defense Technology, Changsha}} 
\newcommand{\IowaState}{\affiliation{Iowa~State~University, Ames, Iowa  50011}}
\newcommand{\ZSU}{\affiliation{Sun Yat-Sen (Zhongshan) University, Guangzhou}}
\newcommand{\CQU}{\affiliation{Chongqing University, Chongqing}} 
\newcommand{\BCC}{\altaffiliation[Now at ]{Department of Chemistry and Chemical Technology, Bronx Community College, Bronx, New York  10453}} 

\newcommand{\UCI}{\affiliation{Department of Physics and Astronomy, University of California, Irvine, California 92697}} 
\newcommand{\GXU}{\affiliation{Guangxi University, No.100 Daxue East Road, Nanning}} 
\newcommand{\HKUST}{\affiliation{The Hong Kong University of Science and Technology, Clear Water Bay, Hong Kong}} 
\newcommand{\Rochester}{\altaffiliation[Now at ]{Department of Physics and Astronomy, University of Rochester, Rochester, New York 14627}} 

\newcommand{\LSU}{\altaffiliation[Now at ]{Department of Physics and Astronomy, Louisiana State University, Baton Rouge, LA 70803}} 

\author{F.~P.~An}\ZSU
\author{W.~D.~Bai}\ZSU
\author{A.~B.~Balantekin}\Wisconsin
\author{M.~Bishai}\BNL
\author{S.~Blyth}\NTU
\author{G.~F.~Cao}\IHEP
\author{J.~Cao}\IHEP
\author{J.~F.~Chang}\IHEP
\author{Y.~Chang}\NUU
\author{H.~S.~Chen}\IHEP
% ...


\collaboration{The Daya Bay Collaboration}\noaffiliation
\date{October 4, 2022}

\begin{abstract}
Reactor neutrino experiments play a crucial role in advancing our knowledge of neutrinos. A precise measurement of reactor electron antineutrino flux and spectrum evolution can be key inputs in improving the knowledge of neutrino mass and mixing as well as reactor nuclear physics and searching for physics beyond the standard model.
In this work, the evolution of the flux and spectrum as a function of the reactor isotopic content is reported in terms of the inverse-beta-decay yield at Daya Bay with 1958 days of data and improved systematic uncertainties.
These measurements are compared with two signature model predictions: the Huber-Mueller model based on the conversion method and the SM2018 model based on the summation method.
The measured average flux and spectrum, as well as their evolution with the \PuN~isotopic fraction, are inconsistent with the predictions of the Huber-Mueller model.
In contrast, the SM2018 model is shown to agree with the average flux and its evolution but fails to describe the energy spectrum.
Altering the predicted IBD spectrum from \PuN~fission does not improve the agreement with the measurement for either model.
The models can be brought into better agreement with the measurements if either the predicted spectrum due to \UF~fission is changed
or the predicted \UF, \UE, \PuN, and \PuO~spectra are changed in equal measure.
\end{abstract}

\maketitle

The detection of reactor electron antineutrinos with the inverse-beta-decay (IBD) process plays a crucial role in advancing our knowledge of neutrinos including the discovery of neutrinos~\cite{Discovery}, establishment of large mixing angle solution of neutrino oscillation~\cite{KamLAND}, and the discovery of non-zero mixing angle $\theta_{13}$~\cite{DYB}. Looking forward, the JUNO experiment requires an accurate knowledge of the reactor neutrino spectrum to determine the neutrino mass ordering~\cite{JUNO}.

For commercial reactors, uranium isotopes are introduced at beginning of a fueling cycle and plutonium isotopes are gradually generated.
Four fission isotopes \UF, \UE, \PuN, and \PuO~account for over the 99.7\% of the antineutrino flux with energy above the IBD detection threshold~\cite{DYBRct}.
A reactor antineutrino prediction, the Huber-Mueller (HM) model~\cite{Huber, Mueller}, is determined by converting cumulative beta spectra to antineutrino spectra for \UF, \PuN, and \PuO~and by summing all involved beta decay branches in databases for \UE.
The average of reactor neutrino flux measurements is only 95\%-96\% of the HM prediction, known as the reactor antineutrino
anomaly (RAA)~\cite{Mention:2011rk, NewFlux, RENOFlux, DCOsc}.
Another anomaly is about the spectrum. The measured neutrino spectrum is poorly described by the HM model, e.g.~a notable ``bump" around 5 MeV~\cite{DybReactor,RENO:2015ksa,DCReactor}.

Together with other experimental anomalies at short-baseline~\cite{LSND, MiniBooNE, GALLEX}, the RAA has motivated a new generation of short-baseline reactor neutrino experiments to search for a sterile neutrino~\cite{DANSS, NEOS, Neutrino-4,PROSPECT2, SoLid,STEREO2,BEST}.
The effect of weak magnetism~\cite{Wang:2017htp}, neutron capture~\cite{Huber:2015ouo}, fission-neutron energy~\cite{Littlejohn:2018hqm} and database inaccuracies~\cite{Hayes:2015yka} on the prediction has been postulated.
In particular, approximately 30\% of the antineutrino flux comes from forbidden decays which can imply an uncertainty
as large as the total flux deficit and the bump~\cite{Forbidden1, Forbidden2,Forbidden3,Forbidden4,Sonzogni:2015aoa}.

Another prediction approach is the summation method, which adds up all related decay branches from databases for all four isotopes.
One such example, the SM2018 calculation~\cite{Pandemonium}, with the latest experimental inputs, predicted a uniformly lower flux from
\UF~than the HM model.
Kopeikin et al.~\cite{Kopeikin} reported the measured ratio between cumulative $\beta$ spectra from \UF~and \PuN~that is also systematically lower than the HM prediction.
Both SM2018 and Kopeikin imply a much smaller discrepancy with neutrino flux measurement than HM.

The most recent results from Daya Bay on the total flux in terms of IBD yield, i.e., the number of antineutrinos per fission multiplied by the IBD cross section~\cite{NewFlux} and evolution of the
spectrum as a function of reactor burnup used a 1230-day data sample~\cite{DYBEvo}.
These results showed that the \UF~yield is about 8\% less than the HM prediction while the
\PuN~yield is consistent with the model.
The latest total and energy differential yields from \UF~and \PuN~with a 1958-day data sample are reported in Ref.~\cite{DYBDecomposition}.
Evolution studies have been performed for the NEOS~\cite{Huber:2016xis} and RENO~\cite{RENORct} experiments.

In this Letter, using the 1958-day data sample taken from December 2011 to August 2017 with the Daya Bay experiment~\cite{DYBP17B}, we report the direct measurement of the total and energy differential IBD yields, $\sigma$ and $\sigma^e$, and their evolution with reactor status with improved systematic uncertainties.
Compared to the unfolded spectra of \UF~and \PuN~\cite{DYBDecomposition},
the measurements in this work do not introduce extra uncertainties from the unfolding method and the theoretical uncertainty of \UE~and \PuO~
which allows a more powerful examination of the combined reactor flux and spectrum prediction of the HM and SM2018 models.

The Daya Bay experiment, equipped with eight antineutrino detectors (ADs), measures the electron antineutrinos from six commercial reactors~\cite{DayaBay:2018,DYBnH,DYBnGd}.
The results in this Letter are based on approximately 3.5 million IBD candidates detected with the four near-site ADs.

The IBD process, $\bar\nu_e + p \rightarrow e^+ + n$, is identified by the prompt-delayed coincidence.
The delayed signal corresponds to the neutron captured on gadolinium.
The prompt energy, $E_{\rm{p}}$, including the kinetic energy of positron and its annihilation gammas,
is related to the antineutrino energy $E_{\nu}\approx E_{\rm p} + 0.78~\rm{MeV}$.
The true $E_{\rm{p}}$ deposit is reconstructed as $E_{\rm{rec}}$.
The reconstructed energy resolution is about 8\% at 1 MeV and
a detector response matrix $M(E_{\rm{rec}},E_{\nu})$ is constructed taking into account all detector effects~\cite{DYBP17B}.
The measured energy spectrum is corrected for the spent-nuclear-fuel contribution and the nonequilibrium contribution~\cite{DYBEvo, DYBDecomposition} for each AD and week, instead of being treated as time independent in the previous analysis~\cite{DYBEvo}.

To measure the IBD yield, a quantity $N_{i}^{dw}$ is calculated for the $d^{\rm th}$ AD and $w^{\rm th}$ week, and $i$ is 5, 8, 9, and 1 for  \UF, \UE, \PuN, and \PuO, respectively~\cite{DYBRct}. It describes the number of fissions of an isotope detected by an AD, and
the definition is
\begin{equation}
 N_{i}^{dw} = \sum_{r=1}^{\rm{6~reactors}} \frac{ N^{\rm{Proton}}_{d} \bar{P}^{\rm{sur}}_{drw} \varepsilon} {4\pi L^{2}_{dr}}\frac{W_{rw} T_{dw}}{\sum_{i}f_{irw}e_{i}} f_{irw},
\label{eq:N^i_dw}
\end{equation}
where $N^{\rm{Proton}}_{d}$ is the number of target protons of the $d^{\rm th}$ AD, $\bar{P}^{\rm{sur}}_{dwr}$ is the average survival probability of reactor electron antineutrinos integrated over energy from the $r^{\rm th}$ reactor to the $d^{\rm th}$ AD calculated under 3-active-neutrino framework in the $w^{\rm th}$ week, $\varepsilon$ is the detection efficiency, $L_{dr}$ is the distance of the AD-reactor pair,
$W_{rw}$ is the thermal power of the $r^{\rm th}$ reactor for the $w^{\rm th}$ week, which is provided by the reactor company,
$T_{dw}$ is the running time of that AD in that week, $f_{irw}$ is the fission fraction of the $i^{\rm th}$ isotope in the $r^{\rm th}$ reactor and $w^{\rm th}$ week, and $e_{i}$ is the energy per fission of the isotope~\cite{EPerFission}.
The effective fission fraction for the $i^{\rm th}$ isotope, $F_i$ ($F_5$, $F_8$, $F_9$, and $F_1$), for that AD and week, $F_{i}^{dw}$, is defined by $F_{i}^{dw}\equiv N_{i}^{dw}/N^{dw}$, in which $N^{dw}=\sum_{i=1}^4 N_{i}^{dw}$.

Data are sorted into 13 groups according
to their effective \PuN~fission fraction $F_{9}^{dw}$, which represents the burnup status of reactors and
 is analogous to the use of $F_{5}^{dw}$~\cite{DYBEvo}.
In this data set, $F_{9}$ ranges from approximately 0.22 to 0.36, and $F_{5}$, correspondingly, ranges from 0.66 to 0.49.
The first group corresponds to $F_9$ between 0.22 and 0.24, due to low statistics, with the additional 12 groups each having a 0.01 interval in $F_9$ from 0.24 to 0.36.
The effective fission fraction of the $g^{\rm th}$ group, $F_i^g$, is calculated as $F_i^g=\sum_{d,w\in g}N_{i}^{dw}/\sum_{d,w\in g}N^{dw}$,
where the information in each AD and week are added together if their $F_{9}^{dw}$'s belong the $g^{\rm th}$ group.
The effective fission fractions averaged over all detectors and time ($\bar F_5$, $\bar F_8$, $\bar F_9$, and $\bar F_1$) are (0.564, 0.076, 0.304, and 0.056).

\begin{figure}[!htbp]
	\includegraphics[width=\columnwidth]{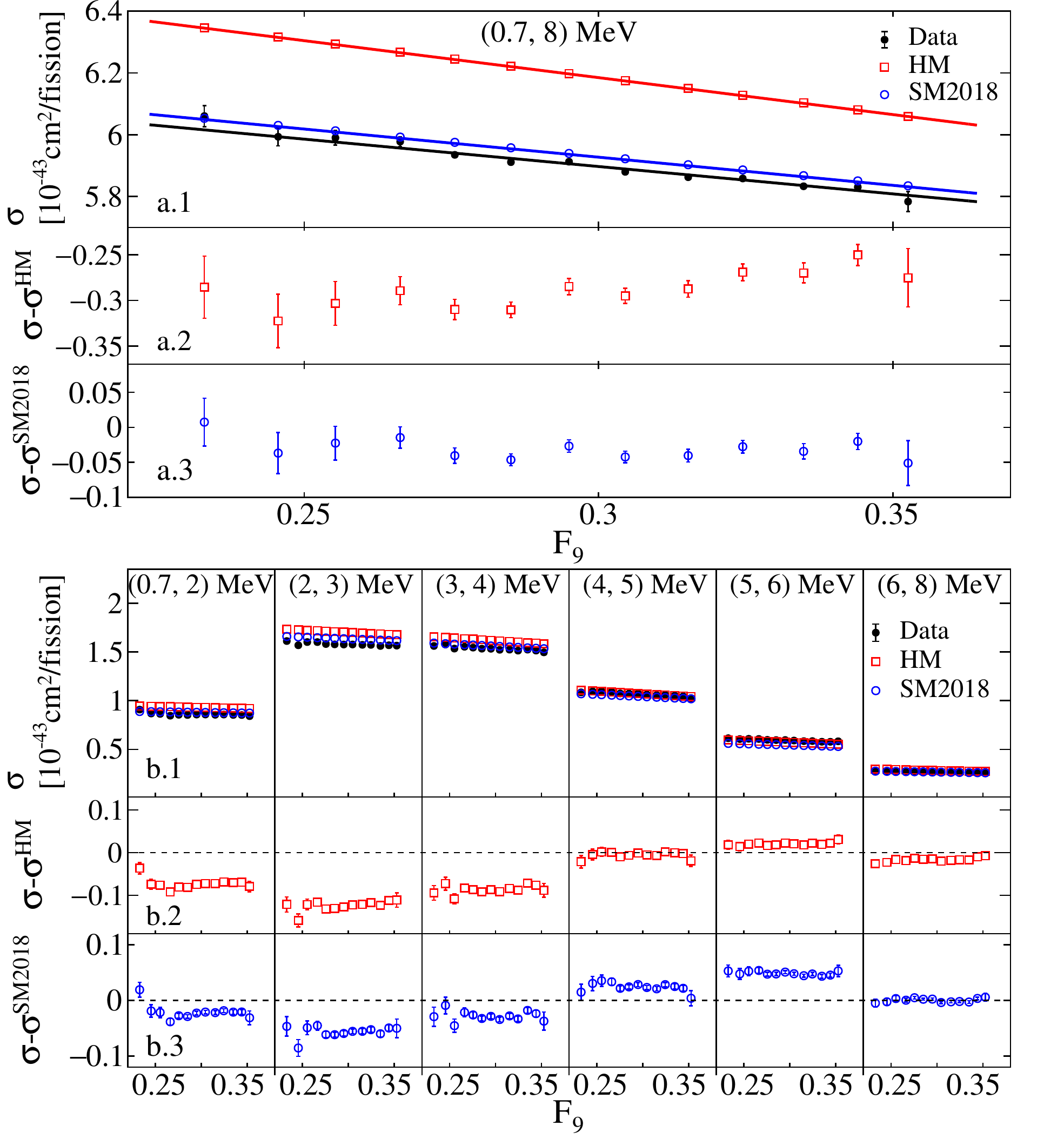}
	\caption{The panels a.1 and b.1 show the total IBD yields in [0.7, 8] MeV and energy differential yield in six reconstructed energy regions as a function of the effective fission fraction of \PuN, $F_9$, respectively.
The best-fit and best-determined lines for the measurements and predictions of the evolution of the total yield are shown in a.1, respectively.
The difference between the measurement and the HM and SM2018 predictions for the total yield (a.2 and a.3) and energy differential yields (b.2 and b.3) are also shown. The error bars represent the statistical uncertainties.
The units of all panels are $10^{-43}~\rm{cm}^2/fission$.}
	\label{fig:Evo}
\end{figure}

The energy differential IBD yield is measured for six reconstructed energy regions: 0.7-2, 2-3, 3-4, 4-5, 5-6, and 6-8~MeV and
the energy differential yield, $\sigma^{eg}$, for the $e^{\rm th}$ energy region and the $g^{\rm th}$ fission group is calculated as~\cite{DYBRct, DYBEvo}
\begin{equation}
\label{eq:NeutrinoYieldMea}
\sigma^{eg}= \int_e \sum_{d,w\in g} S^{dw}(E_{\rm{rec}}) dE_{\rm{rec}} / \sum_{d,w\in g} N^{dwe},
\end{equation}
where the integral is over the energy region,
$S^{dw}(E_{\rm{rec}})$ is the measured energy spectrum of the $d^{\rm th}$ AD in the $w^{\rm th}$ week,
the divisor gives the total number of fissions for the energy region, and the calculation of $N^{dwe}$ is the same $N^{dw}$, except that the neutrino survival probability in Eq.~\ref{eq:N^i_dw} is calculated for the $e^{\rm th}$ $E_{\rm{rec}}$ region only.
The sum over $e$ is the total yield, $\sigma^{g}=\sum_e \sigma^{eg}$, of that group.
The evolution of total and energy differential yield with $F_{9}^{g}$ are plotted in Fig.~\ref{fig:Evo}.

The uncertainties in $\sigma^g$ have statistical, background and the following systematic components.
For the IBD detection efficiency, the AD-correlated uncertainty is improved from 1.7\% to 0.75\%~\cite{NewFlux}, and
the AD-uncorrelated uncertainty is 0.11\%~\cite{DYBDecomposition}.
The uncertainty of the number of target protons is 0.92\% and is AD-correlated~\cite{DYBRct}.
The reactor power measurement uncertainty is 0.5\% and is assigned to be reactor-uncorrelated and time-correlated~\cite{DYBRct}.
The uncertainty of the energy per fission is taken into account~\cite{EPerFission}.
The fission fraction uncertainty for the each isotope and reactor is 5\%, but the uncertainties of the four isotopes are further constrained with the normalization condition and the correlation matrix~\cite{DYBRct} and
are assigned to be reactor- and time-correlated.
The spent nuclear fuel uncertainty is improved from 100\% to 30\%~\cite{DYBP17B}.
The nonequilibrium effect uncertainty is 30\%~\cite{DYBRct}.
The $\theta_{13}$-induced oscillation uncertainty is also included~\cite{DYBP17B}.
The uncertainty of the energy differential yield of $\sigma^{eg}$ further includes
all the energy spectrum uncertainties from the background shape and detector response~\cite{DYBDecomposition},
in which the uncertainties in the absolute energy scale is reduced to be less
than 0.5\% for $E_{\rm{rec}}$ larger than 2 MeV.

The predicted total and energy differential yields of the $i^{\rm th}$ isotope, ($\sigma_{5}$, $\sigma_{9}$, $\sigma_{1}$, and $\sigma_{8}$)
and ($\sigma_{5}^e$, $\sigma_{9}^e$, $\sigma_{1}^e$, and $\sigma_{8}^e$) are obtained
by convolving the product of model prediction and IBD cross section~\cite{DYBRct} with the detector response matrix.
The total yield predictions is defined as
\begin{align}
 \label{eq:HM}
 \sigma^{{\rm Pred},g} \equiv F_{5}^g\sigma_{5}+F_{8}^g\sigma_{8}+F_{9}^g\sigma_{9}+F_{1}^g\sigma_{1},
\end{align}
where $\sigma_i$ are the yields per isotope. Likewise, using the energy differential predictions, $\sigma_i^e$, we define the predicted energy differential yields
\begin{align}
 \label{eq:HMD}
 \sigma^{{\rm Pred},eg} \equiv F_{5}^g\sigma_{5}^e+F_{8}^g\sigma_{8}^e+F_{9}^g\sigma_{9}^e+F_{1}^g\sigma_{1}^e.
\end{align}
The evolution plots of $\sigma^{{\rm{Pred}},g}$ and $\sigma^{{\rm{Pred}},eg}$ with $F_{9}^{g}$ are shown in Fig.~\ref{fig:Evo}.
The differences between the measured and predicted total and energy differential yields are also plotted as a function of $F_{9}^{g}$ in Fig.~\ref{fig:Evo}.

The uncertainties of $\sigma^{{\rm{Pred}},g}$ and $\sigma^{{\rm{Pred}},eg}$ are from all sources involved in the effective fission fraction calculation as described in Eq.~\ref{eq:N^i_dw}, \ref{eq:HM} and \ref{eq:HMD}.
Model uncertainties are poorly defined and not included unless explicitly stated otherwise.

The total yield evolution is compared to the predictions with two characteristic variables, average yield $\bar{\sigma}$ and normalized evolution slope $(d\sigma/dF_{9})/\bar{\sigma}$.
The average yield of $\bar{\sigma}$ and slope of $d\sigma/dF_{9}$ are two direct observables in Fig.~\ref{fig:Evo}.
The evolution of the predicted yield can be described as a linear function of $F_9$ for the observed range of $F_9$.
In addition, if the prediction in Eq.~\ref{eq:HM} is off by a normalization factor $\eta$,
for example, induced by large-mass sterile neutrinos~\cite{Mention:2011rk,DYBSterile,SterileEq} or by a global uncertainty, e.g.~from the detection efficiency,
the prediction would be
\begin{align}
 \label{eq:HM-N}
 \sigma^{{\rm{PredN}},g} = \eta(F_{5}^g\sigma_{5}+F_{8}^g\sigma_{8}+F_{9}^g\sigma_{9}+F_{1}^g\sigma_{1}).
\end{align}
The comparison in the normalized evolution slope $(d\sigma/dF_{9})/\bar{\sigma}$ is free of any normalization issue.

\begin{figure}[!h]
	\includegraphics[width=0.9\columnwidth]{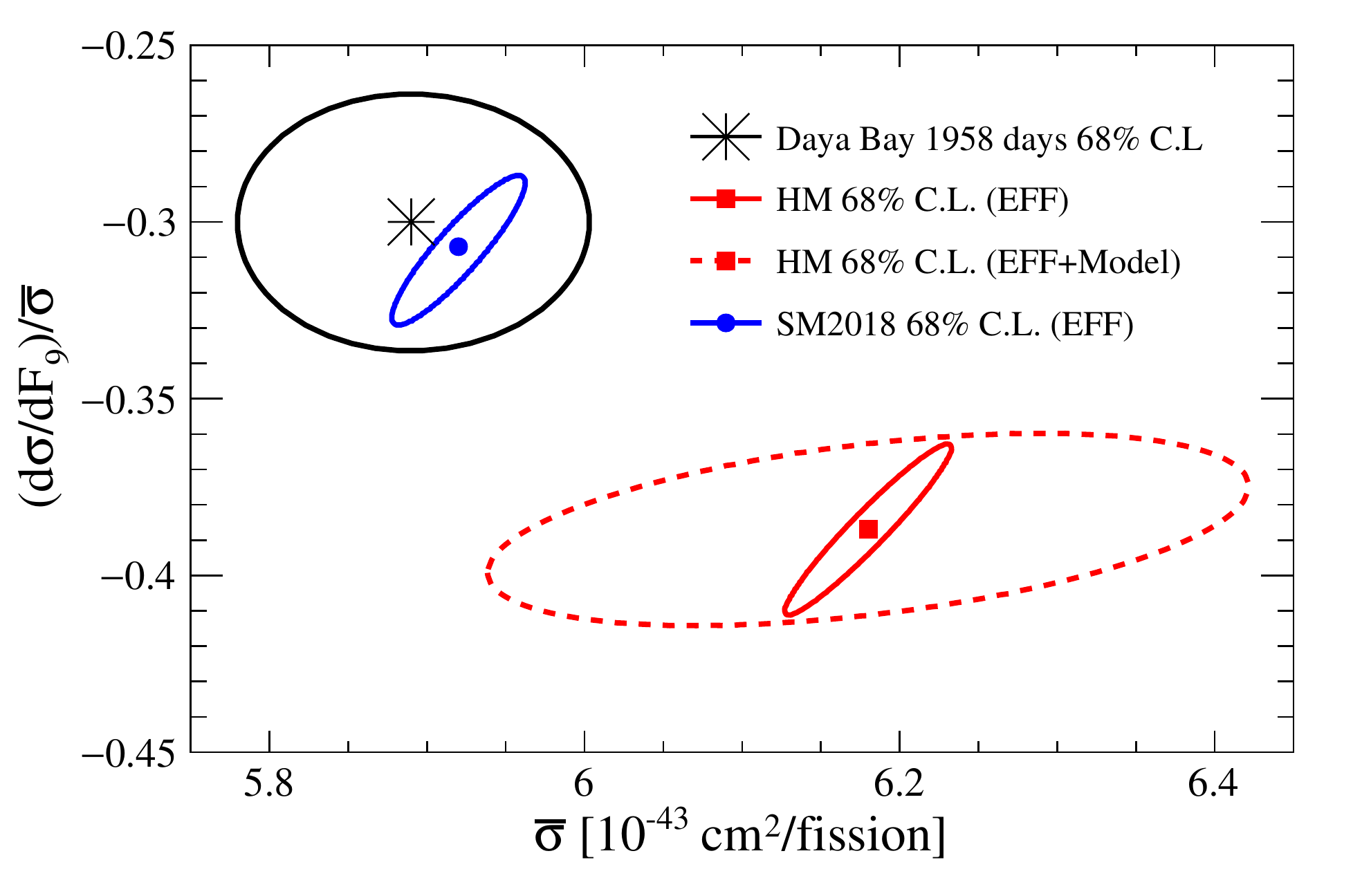}
	\caption{The measured $\bar\sigma$ and $(d\sigma/dF_{9})/\bar{\sigma}$ and their 68\% confidence level (C.L.) contour is shown.
    The predictions of the HM and SM2018 models are shown with their 68\% C.L.~contours with effective fission fraction (EFF) uncertainty.
    The HM model 68\% C.L.~contour including its model uncertainties~\cite{Huber, Mueller} is also shown.}
	\label{fig:EvoContour}
\end{figure}

The total yield measurements in the 13 fission groups
are fitted to the following linear function,
\begin{equation}
\label{eq:LinearFunc}
  \sigma^{{\rm{Lin}},g}=\bar{\sigma}\{1+[(d\sigma/dF_{9})/\bar{\sigma}](F_{9}^{g}-\bar{F}_{9})\},
\end{equation}
with the $\chi^2$ function,
\begin{equation}
\label{eq:LineChi}
  \chi^2=\sum_{gg'} (\sigma^{g}-\sigma^{{\rm{Lin}},g}) (V^{-1})^{gg'} (\sigma^{g'}-\sigma^{{\rm{Lin}},g}),
\end{equation}
to extract $\bar\sigma$ and $(d\sigma/dF_{9})/\bar{\sigma}$,
where $V$ is a $13\times13$ covariance matrix
determined by randomly sampling all the related uncertainty sources described above.
The best-fit results are $\bar\sigma$ = $\AveYieldMeas$ and $[(d\sigma/dF_{9})/\bar{\sigma}]$ = $\DblRatioMeas$
with the $\chi^2$ over the number of degrees of freedom, $\chi^2/\rm{NDF}$, of 9.6/11.
The dominant uncertainty of $\bar\sigma$ is from the IBD detection efficiency and number of target protons.
The dominant uncertainty of $(d\sigma/dF_{9})/\bar{\sigma}$ is from statistics.
The uncertainties from the effective fission fraction calculation are not significant for them.
The best-fit line is shown in Fig.~\ref{fig:Evo}, and the results and 68\% confidence level contour are shown in Fig.~\ref{fig:EvoContour}.

For predictions, $\bar{\sigma}$ and $(d\sigma/dF_{9})/\bar{\sigma}$ can be directly calculated for a set of known fission fractions at Daya Bay.
A joint distribution of $\bar{\sigma}$ and $(d\sigma/dF_{9})/\bar{\sigma}$ is obtained by randomly sampling the effective fission fractions according to their covariance matrix. The mean values and uncertainties of $\bar{\sigma}^{\rm{Pred}}$ and $[(d\sigma/dF_{9})/\bar{\sigma}]^{\rm{Pred}}$ are obtained with the distribution.
The results for the HM are $\bar\sigma^{\rm{HM}}$ = $\AveYieldHM$ and $[(d\sigma/dF_{9})/\bar{\sigma}]^{\rm{HM}}$ = $\DblRatioHM$ ($\AveYieldHMM$ and $\DblRatioHMM$ if including the model uncertainties~\cite{Huber, Mueller}).
The HM prediction in $\bar{\sigma}$ and $(d\sigma/dF_{9})/\bar{\sigma}$ are rejected at 3.6 and 3.0 standard deviations.
For SM2018, the results are consistent with the Daya Bay measurements.
These results are shown in Fig.~\ref{fig:EvoContour} and
the best-determined lines are plotted in Fig.~\ref{fig:Evo}.

The energy differential yield evolution is compared to models with the average yields and normalized evolution slopes in six reconstructed energy regions.
The data are simultaneously fitted to six linear functions,
\begin{equation}
\label{eq:LinearFunc2}
  \sigma^{{\rm{Lin}},eg}=\bar{\sigma}^e\{1+[(d\sigma/dF_{9})/\bar{\sigma}]^e(F_{9}^{g}-\bar{F}_{9})\},
\end{equation}
with the $\chi^2$ function,
\begin{equation}
\label{eq:LineChi2}
  \chi^2=\sum_{ege'g'} (\sigma^{eg}-\sigma^{{\rm{Lin}},eg}) (U^{-1})^{ege'g'} (\sigma^{e'g'}-\sigma^{{\rm{Lin}},e'g'}),
\end{equation}
to extract six pairs of parameters of $\bar{\sigma}^e$ and $[(d\sigma/dF_{9})/\bar{\sigma}]^e$,
where $U$ is a $78\times78$ covariance matrix with a combined row (column) index of $eg$ ($e'g'$)
for the $e^{\rm th}$ ($e'^{\rm th}$) reconstructed energy region and $g^{\rm th}$ ($g'^{\rm th}$) fission fraction group.
$U$ is also determined by a random sampling method of all the related uncertainty sources described earlier.
The best-fit $\chi^2/\rm{NDF}$ is 76/66.
The fit also gives the $12\times12$ covariance matrix of $\bar{\sigma}_e$ and $[(d\sigma/dF_{9})/\bar{\sigma}]^e$,
which includes the $6\times6$ covariance matrix, $A$, for the six $\bar{\sigma}^e$
and the $6\times6$ covariance matrix, $B$, for $[(d\sigma/dF_{9})/\bar{\sigma}]^e$.
The six $\bar{\sigma}^e$ results are strongly correlated because their dominant uncertainties are from the IBD detection
efficiency and number of target protons, and the matrix $A$ deviates strongly from a diagonal matrix.
The six $[(d\sigma/dF_{9})/\bar{\sigma}]^e$ results are all limited by data statistics and largely uncorrelated, and $B$ is close to diagonal.
The correlation between $\bar{\sigma}^e$ and $[(d\sigma/dF_{9})/\bar{\sigma}]^e$ is insignificant.

\begin{figure}[!h]
	\includegraphics[width=\columnwidth]{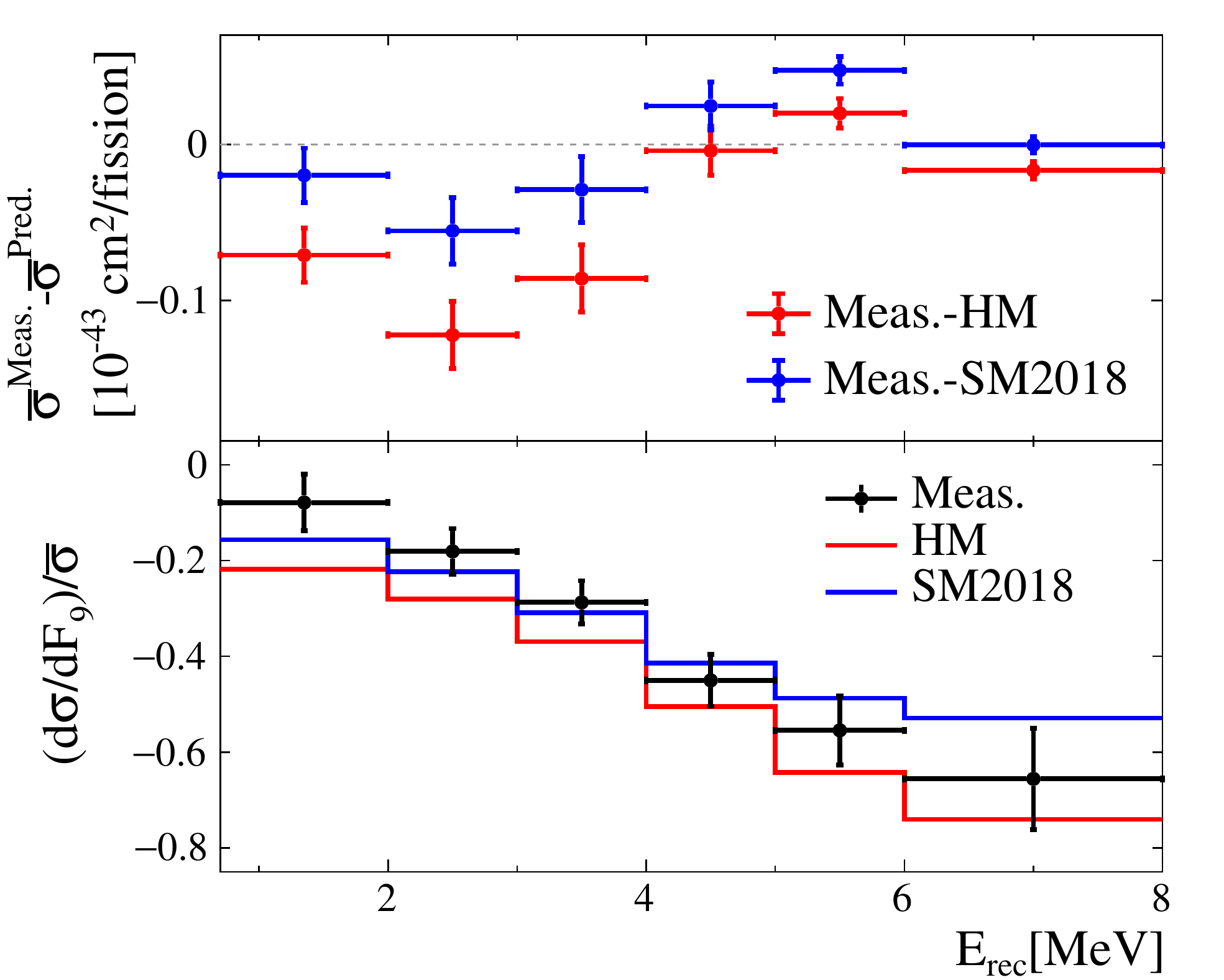}
	\caption{The upper panel shows the difference between the measured energy differential yields and predictions for six reconstructed energy bins, where the error bars are from the measurement. The lower panel shows the normalized evolution slopes for the measurement and predictions, where the uncertainties of measurement are shown.}
	\label{fig:Differential}
\end{figure}

For the predictions, a joint 12-dimension distribution of $\bar{\sigma}^e$ and $[(d\sigma/dF_{9})/\bar{\sigma}]^e$ is obtained by randomly sampling the effective fission fractions as for the study of the predicted total yield and its normalized evolution rate.
The mean values of $\bar{\sigma}^{{\rm{Pred}},e}$ and $[(d\sigma/dF_{9})/\bar{\sigma}]^{{\rm{Pred}},e}$ are obtained with the distribution
as well as the covariance matrix for $\bar{\sigma}^{{\rm{Pred}},e}$, $A^{\rm{Pred}}$, and
the covariance matrix for $[(d\sigma/dF_{9})/\bar{\sigma}]^{{\rm{Pred}},e}$, $B^{\rm{Pred}}$.
The difference of $\bar{\sigma}^{{\rm{Pred}},e}$ with the measurement
and $[(d\sigma/dF_{9})/\bar{\sigma}]^{{\rm{Pred}},e}$ results are plotted in Fig.~\ref{fig:Differential}.
The uncertainty associated with prediction is much smaller than that from measurement.

The average IBD yields of six energy regions, $\bar{\sigma}^e$ are compared to the HM and SM2018 predictions $\bar{\sigma}^{{\rm{Pred}},e}$.
Their difference is quantified as a $\chi^2$ calculated with the difference of $\bar{\sigma}^e$-$\bar{\sigma}^{{\rm{Pred}},e}$ and
their covariance matrix of $A$+$A^{\rm{Pred}}$.
The resulting $\chi^2/\rm{NDF}$ and
the corresponding extent of discrepancy in standard deviations are shown in Tab.~\ref{tab}.
The models do not agree with Daya Bay, and because of
the deficit around 3 MeV and/or the bump around 5 MeV found in the measurement (Fig.~\ref{fig:Evo}) and
the strong correlation among the measurements in different energy regions, their $\chi^2/\rm{NDF}$'s are rather large,
and they correspond to 25 and 27 standard deviations for the HM and SM2018 models, respectively.
The latter, due to the larger discrepancy in the 4-6 MeV region with the measurement, has a slightly worse $\chi^2/\rm{NDF}$ than HM.

The normalized evolution slopes of the six energy regions, $[(d\sigma/dF_{9})/\bar{\sigma}]^e$, are compared to HM and SM2018.
Their difference is quantified with a $\chi^2$ calculated with the difference of $[(d\sigma/dF_{9})/\bar{\sigma}]^e$-$[(d\sigma/dF_{9})/\bar{\sigma}]^{{\rm{Pred}},e}$ and their covariance matrix of
$B$+$B^{\rm{Pred}}$. The resulting $\chi^2/\rm{NDF}$ is shown in Tab.~\ref{tab}.
While the HM and SM2018 models poorly predict the spectral shape,
their predicted relative changes with the fuel composition have much better agreement with the measurement.

\begin{table}
  \centering
  \caption{Comparison results of the measurement with the HM and SM2018 predictions for the average IBD yields of six energy regions, $\bar{\sigma}^e$, (middle column) and the normalized evolution slopes, $[(d\sigma/dF_{9})/\bar{\sigma}]^e$ (right column).
  The $\chi^2/\rm{NDF}$ of the comparison (corresponding number of standard deviations) is given.
  }
  \label{tab}
  \begin{tabular}{c c c}
    \hline
    \hline
    Model        & $\bar{\sigma}^e$ & $[(d\sigma/dF_{9})/\bar{\sigma}]^e$ \\\hline
    HM           & 675/6 (25)  & 11/6 (1.8) \\
    SM2018       & 748/6 (27)  & 5.5/6 (0.7) \\ \hline\hline
  \end{tabular}
\end{table}

To understand the difference between the Daya Bay differential IBD yield evolution and the predictions,
three types of modified models with new free parameters are introduced on top of the HM and SM2018 predictions.

The first modification to each model is to alter only the \UF~energy differential yield prediction in each reconstructed energy region by the fraction $f_5^{e}$ together with the global normalization factor $\eta$, as in Eq.~\ref{eq:HM-N},
\begin{align}
 \label{eq:PH-1}
 \sigma&^{{\rm{model}},eg} \nonumber \\
       &=\eta [F_{5}^{g}\sigma_{5}^{e}(1+f_5^{e})+F_{8}^{g}\sigma_{8}^{e}+F_{9}^{g}\sigma_{9}^{e}+F_{1}^{g}\sigma_{1}^{e}].
 \end{align}
Depending on what the base model is,
the modified models are further labelled as HM+\UF~and SM2018+\UF.
This is motivated by the fact that the majority of the the neutrino flux is due to \UF.

In the second modification to each model, the prediction is
\begin{align}
 \label{eq:PH-2}
 \sigma&^{{\rm{model}},eg} \nonumber \\
       &=\eta [F_{5}^{g}\sigma_{5}^{e}+F_{8}^{g}\sigma_{8}^{e}+F_{9}^{g}\sigma_{9}^{e}(1+f_9^{e})+F_{1}^{g}\sigma_{1}^{e}],
\end{align}
where only the \PuN~energy differential yield predictions in each reconstructed energy region is allowed to change by the fraction $f_9^e$ together with the global normalization factor $\eta$.
The modified models are labelled as HM+\PuN~and SM2018+\PuN~next.
This is motivated given that \PuN~is the second largest contributor to the neutrino flux.

The third modification to each model is to equally scale the predicted spectra of four isotopes in each reconstructed energy region by the fraction $f_{\rm{E}}^e$,
\begin{align}
  \label{eq:PH-3}
 \sigma&^{{\rm{model}},eg}  \nonumber \\
       &=(1+f_{\rm{E}}^e) [F_{5}^{g}\sigma_{5}^{e}+F_{8}^{g}\sigma_{8}^{e}+F_{9}^{g}\sigma_{9}^{e}+F_{1}^{g}\sigma_{1}^{e}].
\end{align}
The motivation is that particular studies~\cite{Xubo:2018eid,Forbidden2} have suggested that all four isotopes may have a common problem in predictions.
They are labelled as HM+Equ and SM2018+Equ.

\begin{figure}[!th]
	\includegraphics[width=\columnwidth]{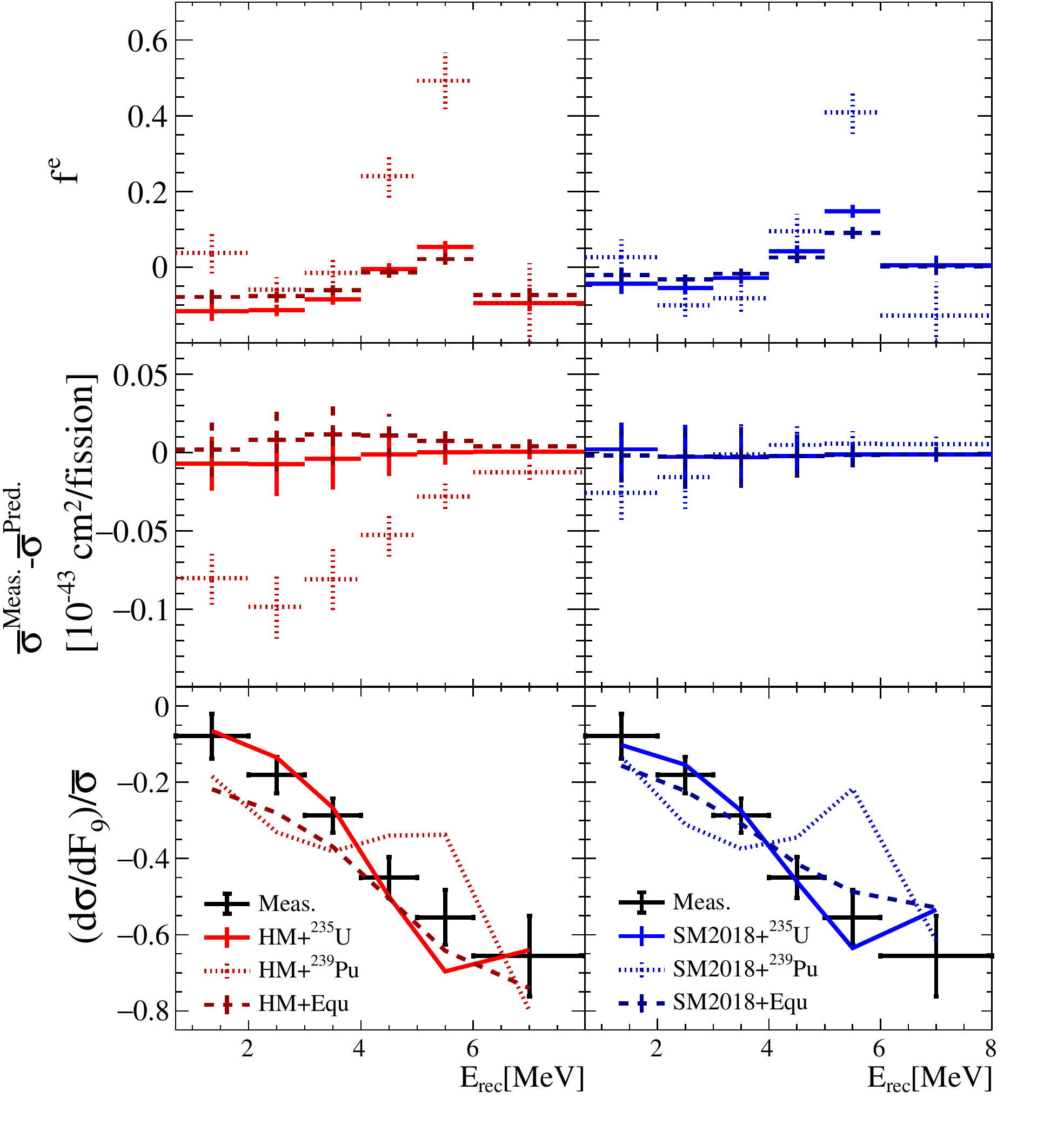}
	\caption{The best-fit $f^e$, i.e.~$f_5^e$, $f_9^e$ or $f_{\rm{E}}^e$, values of the modified models of HM+\UF, SM2018+\UF~(Eq.~\ref{eq:PH-1} with $\eta$ fixed to 1), HM+\PuN, SM2018+\PuN~(Eq.~\ref{eq:PH-2} with $\eta$ fixed to 1), HM+Equ, and SM2018+Equ~(Eq.~\ref{eq:PH-3}) are shown in the upper panels, where the error bars are fit results.
    The deduced $\bar{\sigma}^{{\rm{model}},e}$ predictions with the corresponding $f^e$ values for each model are shown as the difference with the measurement in the middle panels and the error bars shown are from the measurement.
    The measured $[(d\sigma/dF_{9})/\bar{\sigma}]^{e}$ and deduced $[(d\sigma/dF_{9})/\bar{\sigma}]^{{\rm{model}},e}$ are shown in the lower panels and only the error bars of measurement are shown.
    }
	\label{fig:235}
\end{figure}

We fit the measured energy differential yields in the 6 energy regions and 13 fission fraction groups to the modified models with the following $\chi^2$
\begin{equation}
\label{eq:PHChi2}
  \chi^2=\sum_{eg e'g'} (\sigma^{eg}-\sigma^{{\rm{model}},eg}) (Q^{-1})^{ege'g'} (\sigma^{e'g'}-\sigma^{{\rm{model}},e'g'}),
\end{equation}
where six $f_e$'s and/or $\eta$ are free parameters and
$Q$ is a $78\times78$ covariance matrix including all uncertainties for the measurement and predictions determined as $V$ of Eq.~\ref{eq:LineChi} or $U$ of Eq.~\ref{eq:LineChi2}.
When using Eq.~\ref{eq:PH-1} or Eq.~\ref{eq:PH-2}, fits are also performed with $\eta$ fixed to 1.
The best-fit $\chi^2/\rm{NDF}$, the corresponding extent of discrepancy in standard deviations, and best-fit $\eta$, when applicable, are shown in Tab.~\ref{tab:fitresult}.
The best-fit $f_5^e$ and $f_9^e$ of Eq.~\ref{eq:PH-1} and Eq.~\ref{eq:PH-2} with $\eta$ fixed to 1, and $f_E^e$ in Eq.~\ref{eq:PH-3} are shown in Fig.~\ref{fig:235}.
The difference of the deduced $\bar{\sigma}^{{\rm{model}},e}$ with measurement and the deduced $[(d\sigma/dF_{9})/\bar{\sigma}]^{{\rm{model}},e}$ are also shown in the figure, and where the first and third model modifications are preferred with respect to the second model.

\begin{table}[!th]
  \caption{For the six modified models in Eq.~\ref{eq:PH-1},~\ref{eq:PH-2}, and~\ref{eq:PH-3} (the first column),
  the best-fit $\chi^2/\rm{NDF}$ when fitting to data and the corresponding number of standard deviations are shown in the second column
  and the determined normalization factor $\eta$ in the third column.
  Trials are also done with $\eta$ fixed to 1 for Eq.~\ref{eq:PH-1} and Eq.~\ref{eq:PH-2}.}
  \begin{tabular}{ccc}
    \hline
    \hline
      Model & $\chi^{2}$/NDF & $\eta$ \\
    \hline
    {HM+$^{235}$U}     & 83/71 (1.4) & 0.985$\pm$0.021     \\
                       & 83/72 (1.4) & 1 (fixed)   	     \\
    {SM2018+$^{235}$U} & 80/71 (1.2) & 0.997$\pm$0.021    \\
                       & 80/72 (1.2) & 1 (fixed)           \\

    {HM+$^{239}$Pu}    & 116/71 (3.4)& 0.935$\pm$0.014   \\
                       & 136/72 (4.5)& 1 (fixed)           \\
    {SM2018+$^{239}$Pu}& 126/71 (4.0)& 0.995$\pm$0.014   \\
                       & 127/72 (4.0)& 1 (fixed)      	    \\
    HM+Equ             & 89/72 (1.7) & NA \\
    SM2018+Equ         & 82/72 (1.3) & NA \\
     \hline
     \hline
  \end{tabular}
  \label{tab:fitresult}
\end{table}

Even when the \PuN~energy spectra are modified, both the HM and SM2018 model predictions remain incompatible with the data at well over three standard deviations as shown in Tab.~\ref{tab:fitresult}.
For both models, as seen in Fig.~\ref{fig:235}, the required changes of the \PuN~spectrum in some regions are higher than 40\%,
which is far beyond the range of uncertainties caused by the various postulated mechanisms~\cite{Hayes:2015yka,Wang:2017htp,Huber:2015ouo,Littlejohn:2018hqm,Forbidden1,Forbidden2,Forbidden3,Forbidden4,Sonzogni:2015aoa} and is unreasonable.
This observation can be phenomenologically traced back to the features of Fig.~\ref{fig:Evo}.
For example, the $\bar{\sigma}^e$-$\bar{\sigma}^{{\rm{HM}},e}$ in the 2-4 MeV region shows a positive slope and is not proportional to $F_9$,
which contradicts the assumption of pure \PuN-caused anomaly~\cite{Huber:2016xis,RENORct}.

The attempts to adjust the predicted spectrum of \UF~or all spectra in equal measure all lead to good agreement with the data using this metric.
As shown in Tab.~\ref{tab:fitresult}, their best-fit $\eta$ results for \UF-adjusting models are all consistent with 1.
The deduced $\bar{\sigma}^{{\rm{model}},e}$ and $[(d\sigma/dF_{9})/\bar{\sigma}]^{{\rm{model}},e}$
are consistent with the measurements as shown in Fig.~\ref{fig:235}.
HM+$^{235}$U works slightly better than HM+Equ model, as their best-fit $\chi^2/\rm{NDF}$ shown in Tab.~\ref{tab:fitresult}.
But with the current precision of the Daya Bay data set, it is difficult to distinguish whether \UF, by itself, or a mix of fission isotopes, are responsible for the flux and spectrum anomalies.

In summary, the total and energy differential IBD yield evolution as a function of fuel composition are measured and compared to the predictions of two signature models: the HM model based on the conversion method and the SM2018 model based on the summation method.
While the measurement of the total IBD yield evolution is found to be incompatible with the HM model prediction, it is consistent with the SM2018 prediction.
On the other hand, the predictions of spectrum evolution for both HM and SM2018 model show large discrepancies from the data.
We exclude at high significance the hypothesis that the \PuN~energy spectrum in HM or SM2018 models is responsible for the entire difference with the data, regardless of how the normalization of the Daya Bay data is treated.
In contrast, good consistency with the data can be achieved either by altering the \UF~spectrum or all four isotopes' spectra in equal measure in the SM2018 model. For the HM model, the \UF~spectrum adjustment works slightly better than adjusting all spectra, as indicated by the total yield evolution measurement. Future enhancements to the models could prioritize \UF-specific causes or factors common to the four isotopes.

Daya Bay is supported in part by the Ministry of Science and Technology of China,
the U.S. Department of Energy,
the Chinese Academy of Sciences,
the National Natural Science Foundation of China,
the Guangdong provincial government,
the Shenzhen municipal government,
the China General Nuclear Power Group,
Key Laboratory of Particle and Radiation Imaging (Tsinghua University), the Ministry of Education,
Key Laboratory of Particle Physics and Particle Irradiation (Shandong University), the Ministry of Education,
Shanghai Laboratory for Particle Physics and Cosmology,
the Research Grants Council of the Hong Kong Special Administrative Region of China,
the University Development Fund of The University of Hong Kong,
the MOE program for Research of Excellence at National Taiwan University,
National Chiao-Tung University, and NSC fund support from Taiwan,
the U.S. National Science Foundation,
the Alfred~P.~Sloan Foundation,
the Charles University Research Center UNCE/SCI/013 in the Czech Republic,
the Joint Institute of Nuclear Research in Dubna, Russia,
the CNFC-RFBR joint research program, the National Commission of Scientific and Technological Research of Chile,
and the Tsinghua University Initiative Scientific Research Program.
We acknowledge Yellow River Engineering Consulting Co., Ltd., and China Railway 15th Bureau Group Co., Ltd., for building the underground laboratory. We are grateful for the ongoing cooperation from the China General Nuclear Power Group and China Light and Power Company.

\bibliographystyle{apsrev4-1}

\end{document}